\documentclass[superscriptaddress,pre,twocolumn]{revtex4}
\usepackage{amsmath,amssymb}
\usepackage{bm}
\usepackage{graphicx,color,natbib}
\usepackage{placeins}
\usepackage[dvipsnames]{xcolor}

\begin{document}
\title{Tuning Higher Order Structure in Colloidal Fluids}

\author{Xiaoyue Wu}
\affiliation{School of Chemistry, University of Leeds, Woodhouse Lane, Leeds, LS2 9JT, United Kingdom.}

\author{Katherine Skipper}
\affiliation{H. H. Wills Physics Laboratory, University of Bristol, Bristol BS8 1TL, United Kingdom.}

\author{Yushi Yang}
\affiliation{H. H. Wills Physics Laboratory, University of Bristol, Bristol BS8 1TL, United Kingdom.}

\author{Fergus J. Moore}
\affiliation{H. H. Wills Physics Laboratory, University of Bristol, Bristol BS8 1TL, United Kingdom.}

\author{Fiona C. Meldrum}
\affiliation{School of Chemistry, University of Leeds, Woodhouse Lane, Leeds, LS2 9JT, United Kingdom.}

\author{C. Patrick Royall}
\affiliation{Gulliver UMR CNRS 7083, ESPCI Paris, Universit\'{e} PSL, 75005 Paris, France.}
\email{paddy.royall@espci.fr}

\date{\today}

\begin{abstract}
Colloidal particles self assemble into a wide range of structures under external AC electric fields due to induced dipolar interactions [Yethiraj and Van Blaaderen \emph{Nature} \textbf{421} 513 (2003)]. As a result of these dipolar interactions, at low volume fraction the system is modulated between a hard--sphere like state (in the case of zero applied field) and a ``string fluid'' upon application of the field. Using both particle--resolved experiments and computer simulations, we investigate the emergence of the string fluid with a variety of structural measures including two-body and higher--order correlations. The higher--order structure we probe using three-body spatial correlation functions and a many--body approach based on minimum energy clusters of a dipolar--Lennard--Jones system. The latter constitutes a series of geometrically distinct minimum energy clusters upon increasing the strength of the dipolar interaction, which are echoed in the higher--order structure of the colloidal fluids we study here. We find good agreement between experiment and simulation at the two-body level. Higher--order correlations exhibit reasonable agreement between experiment and simulation, again with more discrepancy at higher field strength for three--body correlation functions. At higher field strength, the cluster population in our experiments and simulations is dominated by the minimum energy clusters for all sizes $8 \leq m \leq 12$. 
\end{abstract}

\maketitle

\section{Introduction}
\label{sectionIntroduction}

Particles with a dipolar interaction are of significant fundamental importance in the study of fluids and disordered materials. They are among the simplest models which describe long range directional interactions, which are exhibited by molecules~\cite{israelachvili,gebbie2017}. Colloidal dispersions provide suitable models of atomistic and molecular systems as they exhibit phase behaviour following the same rules of statistical mechanics, yet are amenable to real space observation using optical microscopy~\cite{evans2019,ivlev,bharti2015}.

Colloidal dipolar systems fall broadly into two categories. Some, for example ferromagnetic nanoparticle systems, like atoms and molecules, have an intrinsic dipole moment~\cite{boles2016} and can be modeled with the Stockmayer model which combines a dipolar interaction with a Lennard--Jones interaction~\cite{novak2019}. These systems exhibit intriguing string--like structures~\cite{klokkenburg2006,pyanzina2007}, with branching ~\cite{kantorovich2015}, coiling ~\cite{spataforasalazar2023} and clustering ~\cite{novak2019} behavior, not to mention a ferromagnetic transition~\cite{weis2006}. Rather than spontaneous dipolar interactions, in other colloidal systems, dipoles may be induced by an external electric or magnetic field~\cite{ivlev,bharti2015,donaldson2017}. This has the consequence that the dipolar interactions are aligned in the direction of the applied field. Using ferromagnetic and superparamagnetic nanoparticles in an external field then opens further possibilities such as a very strong response to the field~\cite{mostarac2023}. Other, more exotic possibilities include the use of a biaxial field, leading to phenomena such as in--plane condensation in (quasi) 2d systems~\cite{ivlev,elsner2009}, control of assembly and tuning of interactions ~\cite{yakovlev2017,yakovlev2021} and direct observation of cluster growth~\cite{bensalah2023}. In addition to their fundamental interest, such dipolar colloidal systems may find application as electrorheological fluids~\cite{winslow2000,dassanayake2000}, hydraulic valves~\cite{parthasarathy1996} and photonic materials ~\cite{vanblaaderen2004}.

Here we shall focus on dipolar systems with an external field. A particular attraction of these systems is the ease with which the dipolar interactions can be tuned with the external field. Indeed a combination of real~\cite{yethiraj2003} and reciprocal space~\cite{klokkenburg2007,wiedenmann2008} studies of such systems enables  the investigation of a variety of crystal structures including fcc, hcp, bcc, body-centered tetragonal (bct) and body-centered orthorhombic (bco) structures~\cite{yethiraj2003,hynninen2005,bharti2015}, along with a transient labyrinthine structure~\cite{dassanayake2000}.  Tuning the electric (or magnetic) field \emph{in--situ} enables the control of phenomena such as a martensitic transition~\cite{yethiraj2004}. Adding softness~\cite{colla2018} or attractions~\cite{semwal2022} to the interaction potential further increases the range of structures into which the system may self--assemble.

In addition to the rich crystalline phase behavior, dipolar colloids feature a fluid phase at lower colloid volume fraction and electric field strength than those at which the crystals are found. Interestingly, the symmetry--breaking dipolar interactions cause this fluid to assemble into string--like structures which are aligned in the direction of the electric field~\cite{dassanayake2000,yethiraj2003,li2010}. This ``string fluid'' has been investigated analytically~\cite{elfimova2012} and can form the basis for producing ``colloidal polymers''~\cite{vutukuri2012,wei2016}. Meanwhile, it is possible to take the system out of equilibrium, which enables investigation of string growth mechanisms~\cite{bensalah2023}, and aggregation phenomena between the strings~\cite{messina2023}. This suggests that the structure of the ``string fluid'' may be rather interesting and it has been investigated analytically~\cite{elfimova2012} and in reciprocal space~\cite{pyanzina2007} in addition to real space~\cite{yethiraj2003,semwal2022}.

In atomic and molecular systems, characterizing structure in the fluid state beyond pair correlations is challenging, although not impossible~\cite{royall2015physrep,zaluzhnyy2019}. With colloidal systems, particle--resolved studies~\cite{ivlev} which deliver coordinates in real space are amenable to the measurement of higher--order correlation functions, relevant to a variety of phenomena such as dynamical arrest~\cite{vanblaaderen1995,royall2008,leocmach2013,royall2015physrep} and polymorph selection~\cite{taffs2016,leoni2021,gispen2023} and crystal precursors ~\cite{russo2012,leoni2021}. Furthermore, theoretical treatments have been developed to describe the higher--order structure of hard sphere colloids~\cite{hansen,robinson2019prl,robinson2019pre}.

Methods to characterize higher--order structure include three--body correlation functions such as $g_3$~\cite{hansen,russ2003,royall2015prl}, and higher--order correlations such as common neighbor analysis (CNA)~\cite{honeycutt1987} and Voronoi face analysis~\cite{tanemura1977}. These have been shown to be successful in studying the structure of fluids and glasses~\cite{jonsson1988,bailey2004}. Another strategy, the bond orientation order (BOO) parameters developed by Steinhardt \emph{et al.} \author{steinhardt1983}~\cite{steinhardt1983}, focuses on the local symmetry around a central particle. This method has been shown to be very useful in the study of crystallization, especially in identification of small crystalline clusters in a supercooled liquid~\cite{auer2004,kawasaki2010} and also in the characterization of fivefold symmetric order in amorphous systems~\cite{vanblaaderen1995}.
With the popularity of machine learning rising, it has been applied effectively to local structure in amorphous materials, for example by combining many structural metrics such as the pair correlation function~\cite{cubuk2015,leoni2021}. Other examples include combining it with local descriptors such as CNA and BOO to better characterize the local environment around a single particle in disordered materials~\cite{boattini2020,boattini2019jcp,leoni2021}. Both supervised~\cite{gerardo2021} and unsupervised learning~\cite{joris2020} have been used to further our understanding of supercooled liquid and glass forming systems.

The methods discussed above are geometric in nature. An alternative approach, which takes into account the interactions between the constituent particles of the system, has its roots in the work of Sir Charles Frank~\cite{frank1952}, who postulated that since the minimum potential energy configuration of 13 Lennard-Jones atoms corresponds to an icosahedron, that this would be a common geometric motif in (supercooled) liquids. With the advent of energy landscape calculations~\cite{wales}, it has become possible to determine the structure of minimum potential energy clusters for a wide range and size of systems including the Lennard-Jones~\cite{wales1997}, ``Sticky Sphere'' ~\cite{trombach2018}, Stockmayer~\cite{miller2005} models and more exotic interactions relating to multiple fields~\cite{miller2008}. Since the dipolar interaction of the Stockmayer model is not constrained to lie in any particular direction it thus corresponds to a molecular (or nanoparticle~\cite{novak2019}) system, rather than a colloidal dipolar system in an external field in the context of the discussion above.

\begin{figure}
\includegraphics[width=85mm]{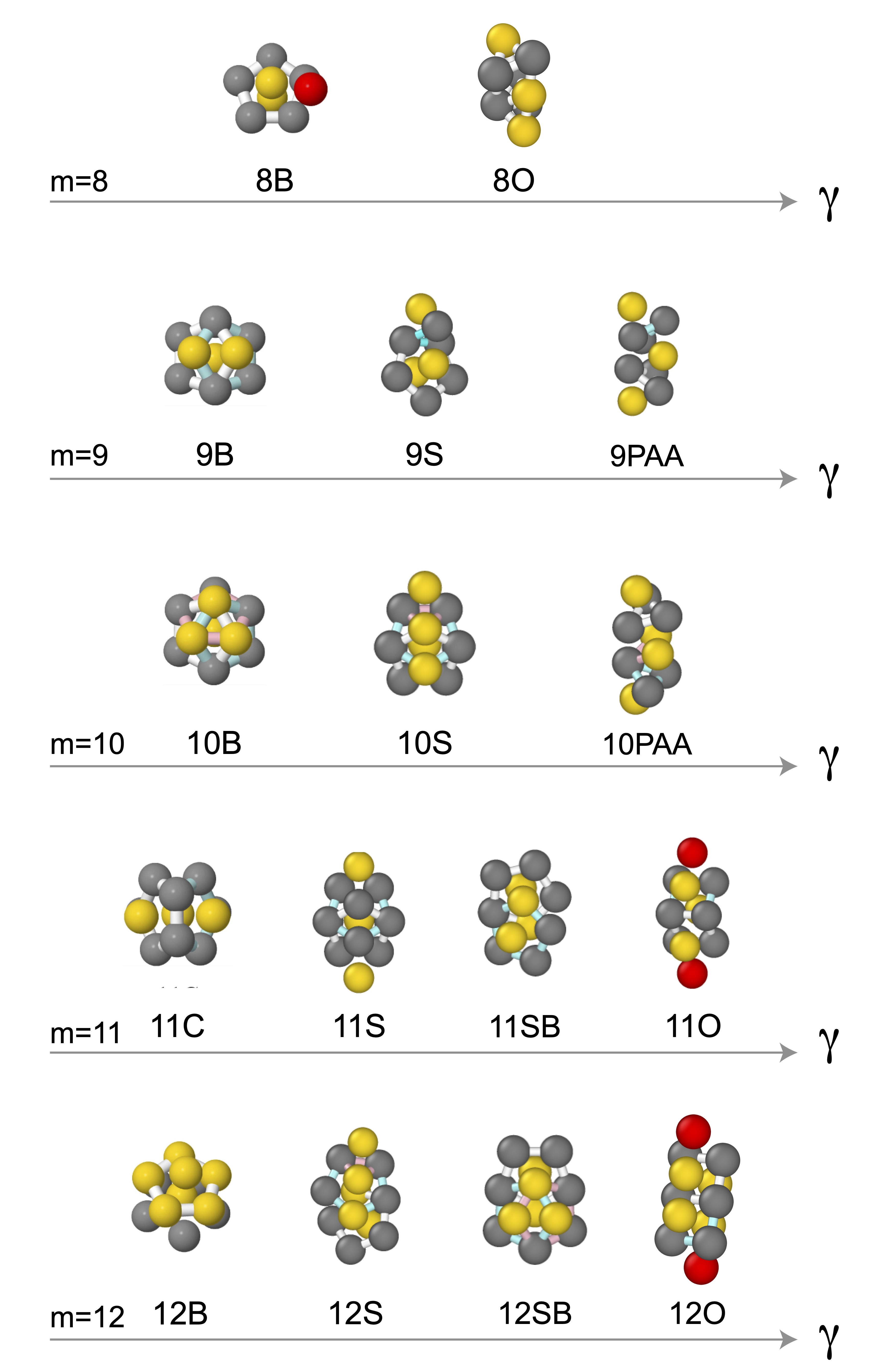}
\caption{Rigid minimum energy clusters of the dipolar--Lennard-Jones system for various sizes $m$.
8B, 9B, 10B, 11C and 12B are minimum energy clusters for the Lennard--Jones system ($\gamma=0$). 
Different geometries correspond to minimum energy clusters as a function of of the dipolar strength $\gamma$. Here we consider rigid clusters only. The clusters are formed from rings of three, four or five particles. These are coloured grey. In the axis perpendicular to the rings are so--called spindle particles, colored yellow, one above and one below the ring. Single spindle particles are coloured red~\cite{malins2013tcc,skipper2024}.}
\label{figTCCClusters}
\end{figure}

Identifying local arrangements of particles in bulk systems whose bond network is identical to such clusters can be carried out using the \emph{topological cluster classification} (TCC)~\cite{malins2013tcc,skipper2024}. The TCC has been used to identify locally favored structures or minimum energy clusters in systems undergoing dynamic arrest~\cite{royall2008}, colloid-polymer mixtures interacting via the Morse potential~\cite{taffs2010jpcm,klix2013}, colloidal suspensions with attractive interactions~\cite{taffs2010jcp}, colloidal gels~\cite{royall2015prl,malins2010} and the liquid--gas interface~\cite{godonoga2010}. However, thus far this method has only been applied to systems with a spherically symmetric interaction, and those whose state point is fixed by the composition of the system. Some of us have recently determined the minimum energy clusters for a Lennard--Jones--dipolar system where the dipoles are induced in a particular direction (Fig.~\ref{figTCCClusters}) ~\cite{skipper2024}, which opens the possibility to use this method to probe the higher--order structure of dipolar colloids. in this way, we can apply the TCC method to an experimental system with asymmetric interactions which can be tuned \emph{in-situ}.

Herein, we report a combined experimental and computer simulation study which carries out such a study of higher-order structure of dipolar colloids. Since the electric field can be tuned at will, it is possible to vary the state point of the system \emph{in situ}. This is somewhat unusual for colloidal systems, where the state point is often fixed by the composition of the system. Here we explore the equilibrium string fluid phase, but we can also increase the field such that the system becomes metastable to a phase coexistence between a fluid and a body-centered tetragonal crystal~\cite{hynninen2005}. We consider pair correlations in the form of radial distribution functions $g_2(r)$ and three-body correlations in the form of order parameters to determine ``string-like'' configurations and also the triplet correlation function $g_3(r,r',\eta)$. We use the topological cluster classification~\cite{malins2013tcc,skipper2024} to explore higher--order spatial correlations in the form of minimum energy clusters of the dipolar--Lennard--Jones interaction~\cite{skipper2024}. Before concluding, we provide an outlook on the study of non-equilibrium behaviour accessed via higher-order correlations.

\section{Dipolar Interactions in Colloidal Systems}

The colloids in our experiments are suspended in an index-matching solvent with added salt, as described in Sec:  \ref{sectionExperimental}. Like very many colloidal systems~\cite{royall2013myth}, ours carry an electrostatic charge. This we screen by adding salt, and we have previously demonstrated that this is sufficient to treat the particles as hard spheres, with a slight increase in effective diameter due to the residual electrostatics~\cite{royall2018jcp}.

When colloids are subjected to an external electric field $\mathbf{E}$, a dipole-dipole interaction is induced between the particles which reads

\begin{equation}
\frac{u_\mathrm{dip}(r,\theta)}{k_BT}= \frac{\gamma}{2} \left(\frac{\sigma}{r}\right)^3 \left(1-3 \cos^2 \theta \right)
\label{eqUDipole}
\end{equation}

\noindent
where $u_\mathrm{dip}$ is the dipolar interaction, $k_BT$ is the thermal energy and $\theta$ is the angle made by $\mathbf{r}$ and the $z$-axis. 
Here $\mathbf{r}$ is the vector between the centres of the colloids. In our experimental system, $\gamma=\gamma_\mathrm{exp}$ is a dimensionless prefactor that depends on the strength of the external field and material properties of the system. Here $\mathbf{r}$ is the vector connecting the centres of the two particles. The prefactor

\begin{equation}
\gamma_\mathrm{exp}=\frac{\mathbf{p}^2} {2\pi \varepsilon_s \varepsilon_0 \sigma^3 k_BT}
\label{eqGamma}
\end{equation}

\noindent
where $\varepsilon_s$ is the dielectric constant of the solvent and the dipolar moment

\begin{equation}
\mathbf{p}=\frac{\pi}{2} \alpha \varepsilon_s \varepsilon_0 \sigma^3 \mathbf{E}_\mathrm{loc}
\label{eqDipoleMoment}
\end{equation}

\noindent
Here $\alpha = (\varepsilon_p - \varepsilon_s) / (\varepsilon_p + 2\varepsilon_s)$ where
$\varepsilon_p$ and $\varepsilon_s$ is the dielectric constant of the particles and solvent respectively and $\mathbf{E}_\mathrm{loc}$ is the local electric field. Here we follow~\citet{hynninen2005} and set $\mathbf{E}_\mathrm{loc}=\mathbf{E}/(1- \alpha\pi /6)$ which is the case for a cubic crystal.

Combining with the hard sphere interaction noted above $u_\mathrm{hs}$, the total interaction between two colloids under the electric field becomes

\begin{equation}
u_\mathrm{total}(r,\theta) = u_\mathrm{hs}(r) + u_\mathrm{dip}(r,\theta).
\label{eqUTotal}
\end{equation}

\section{Methods}

\subsection{Experimental}
\label{sectionExperimental}

The colloidal suspension used in this experiment was prepared by adding sterically-stabilised polymethyl methacrylate (PMMA) spheres (synthesized following reference~\cite{campbell2002,hollingsworth2006}) ($\rho$ = 1.196 gcm$^{-3}$)~\cite{royall2005s} of diameter $\sigma=1.73$ $\mu$m 
and polydispersity $\lesssim 5\%$~\cite{donovan} 
~\footnote{A value of 3.8\% was obtained for the polydispersity using static light scattering, and a higher value was obtained using electron microscopy. Different methods are known to provide different results when measuring the size polydispersity~\cite{poon2012,royall2023}. In this case, that the particles crystallize readily suggests that their polydispersity was around 5\% or less.}
to a mixture of density and refractive index matched solvents. The particles we labelled with 1,1'-dioctadecyl-3,3,3',3'-tetramethyl-indocarbocyanine perchlorate, which may be excited at a wavelength of 543 nm~\cite{campbell2002}. The solvent is a mixture of \emph{cis}-decalin ($\rho \approx $ 0.897 gcm$^{-3}$) and cyclohexyl-Bromide (CHB) ($\rho$ = 1.32 gcm$^{-3}$). Tetrabutylammonium bromide (TBAB) salt was dissolved in the solvent to make up a solution with TBAB concentration of 260 $\mu$M. This corresponds to a Debye length $\kappa^{-1}$ of around 100 nm~\cite{taffs2013}.  Since the Debye length is much less than the particle diameter, in the absence of an electric field, the colloids behave as nearly hard spheres~\cite{taffs2013,royall2013myth}. While more sophisticated treatments may be carried out to match the interaction potential~\cite{taffs2013,royall2013myth}, here we use a slightly soft potential in the computer simulations and presume this to be sufficient to match the experimental system, noting that the effects we seek to study are dominated by the dipolar interactions, rather than the hard core or precise value of the colloid volume fraction (which we determine by weighing out the samples) ~\cite{poon2012}.

We determine the dipolar contribution to the interaction potential between the particles by evaluating Eqs.~\ref{eqGamma} and ~\ref{eqDipoleMoment} with the particle diameter $\sigma$, the solvent dielectric constant $\epsilon_m=5.6$~\cite{leunissenThesis} and the measured value of the local electric field $\mathbf{E}$. We emphasize that the resulting values of $\gamma_\mathrm{exp}$ have no fit parameters and are purely dependent on the material properties of the system. See Sec.~\ref{sectionDiscussion} for further discussion as to the importance of the absence of fit parameters.

In order to construct the sample cell to hold the colloidal suspension, two indium tin oxide glass slides were separated with spacer silica particles of approximately 60 $\mu$m in diameter. This created a transparent sample cell  with electrodes top and bottom.  The electrodes were connected to a signal generator to provide an AC electric field across the cell.

A Leica SP8 confocal microscope was used to monitor the system under the applied electric field. During each measurement, a stack of 3d confocal images of at least 200 ``slices'' of $xy$ images along the $z$ direction were taken with 256$\times$256 pixels with at least ten pixels per particle diameter in all directions. 3d images were acquired at least 30s apart. Following related work with dipolar colloids~\cite{yethiraj2003,leunissenThesis}, only particles at least ten diameters from the wall were analyzed, to ensure that there were no significant wall effects. We saw no influence from the wall proximity in any of our measurements and we conclude that we can treat the system as bulk, as is typical for such particle--resolved studies. It is worth noting that the system sizes used in this experimental technique are not huge~\cite{ivlev}. The shortest dimension of the system (here 60 $\mu$m) is two orders of magnitude larger than that of the colloidal particles.

In each measurement, the applied voltage and the thickness of the electrical cell was measured in order to allow electric field strength comparison across different experiments. 
The frequency of the electric field applied was 1Mhz and the field strength was up to 0.3 $V\mu^{-1}$ measured peak-to-peak. Before each measurement, the system is allowed to stabilise for at least 20 min. The Brownian time $\tau_B=(3 \pi \nu \sigma^3)/(4 k_B T) \approx 6.09$ s for our system, so we consider this time quite sufficient to relax equilibrium states. Here $\nu$ is the solvent viscosity. At least 50 3d images each separated by 30s were taken for each state point, in the same place in the sample. In equilibrium, this provides sufficient statistics. However, out of equilibrium, for high field strengths, there will likely be some dependence on the history of the system, to which we return below in Secs.~\ref{sectionResults} and ~\ref{sectionDiscussion}.

\subsection{Particle Tracking}

Here we use a slight modification to enhance the accuracy of the coordinates that we detect~\cite{statt2016}. We begin by carrying out a conventional centroid location~\cite{crocker1995,leocmach2013sm}. This seeks the brightest pixels and weights the brightnesses of the surrounding pixels to obtain an estimate of the centre of the colloidal particle. Overlaps corresponding to multiple pixels within a single particle being identified are removed. This method works well in (quasi) 2d studies~\cite{royall2023}, but in the case of the 3d confocal microscopy that we carry out here, overlaps between blurred images of particles in the $z$-direction can be a problem.

To mitigate such blurring in the $z$--direction, we refine the first set of coordinates determined as described above as follows. Knowing the size of the particles, the algorithm predicts an image based on the set of particle coordinates. This predicted image is then iteratively compared with the original measured image and the coordinates moved following a Monte Carlo method using the difference in pixel values between the predicted and measured image to minimize the differences between them. Further details of our method, including the source code may be found in ~\citet{yang2021}.

Colloid tracking is subject to errors in the location of the coordinates of the particles. Combined with polydispersity in the particle size distribution, this can influence structural measurements as we carry out here. In the case of 2-body correlation functions, the effect of polydispersity and tracking errors can be similar to a convolution~\cite{royall2007jcp}. In the case of amorphous systems, the effect of (mild) polydispersity has been investigated and this was found to have only a minor effect on the higher-order structure~\cite{royall2012}. In Fig. ~\ref{sFigGError} in the appendix, we show the effects of adding a range of tracking errors to coordinates of computer simulation data. The height of the peaks and their width broadens as the error is increased.

\begin{figure}
\includegraphics[width=\linewidth]{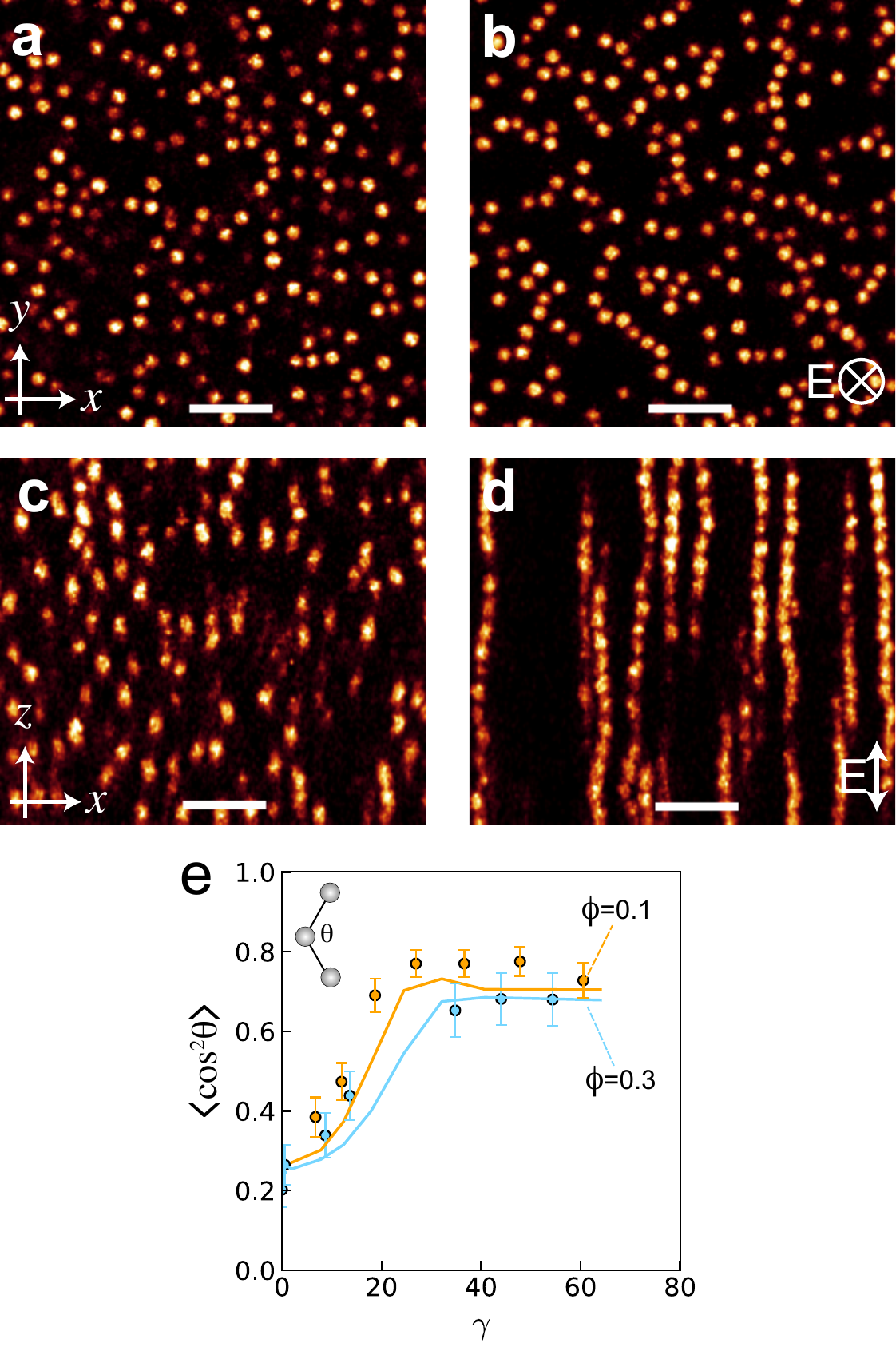}
\caption{The colloidal dipolar system. 
(a-d) Representative images using confocal microscopy in the horizontal $xy$ (a,b) and vertical $xz$ (c,d). Here volume fraction $\phi$=0.1.
(a,c) No field, $\mathbf{E}=0$. 
(b,d) Confocal microscopy images of a system at the same volume fraction at the maximum electric field strength corresponding to $\gamma=60$.
(e) String fluid order parameter ($\langle {\cos^2{\theta}} \rangle$) as a function of external electric field strength $E$. The inset indicates the angle $\theta$.
Data are shown for experiments (data points) and simulations (lines) for volume fractions $\phi=0.1$ and $0.3$ as indicated.
Scale bars in (a-d) are 20 $\mu$m. Error bars are the standard error.
}
\label{figBopString}
\end{figure}

\subsection{Computer Simulation}

We employed computer simulations using the LAMMPS package in the NVT ensemble~\cite{plimpton1995}. To mimick the colloidal motion, we used Langevin dynamics, where the $i$th particle has position $\mathbf{r}_i$ and velocity $\mathbf{v}_i$ and the velocities evolve in time as

\begin{equation}
m\frac{d}{dt} \mathbf{v}_i  = - \nabla_i  V - \mu  \mathbf{v}_i + \sqrt{2 \mu k_B T} \xi_i
\label{eqLangevin}
\end{equation}

\noindent
where $m$ is the mass of the particle, $V$ is the total potential energy 
while $\mu$ is a friction constant and $\xi_i$ a random noise force. The friction constant sets the time scale for the decay of velocity correlations as $t_d = m/\mu$. For colloids, the physical situation corresponds to an overdamped limit with  $t_d$ very small. It has been found that setting $t_d = 0.1$ here is appropriate to following the dynamics of non-equilibrium colloidal systems and can robustly be compared with experiments~\cite{razali2017,zia2014}.

To reproduce the (small) polydispersity in the experimental system, we follow some previous work and use an equimolar five-component system~\cite{royall2018jcp}. Here we set the diameters of the particles as $\sigma_{1\leq i \leq 5} = (0.93675445, 0.968377225, 1, 1.031622775, 1.06324555)$. We do not treat the effect of polydispersity on the dipolar interactions, other than rescaling the interaction range with the mean of the particle diameters. To our knowledge, no such study has been undertaken. We therefore implement the effects of polydispersity through the hard core contribution which we treat as follows.

To reproduce the (nearly) hard sphere behaviour of the experimental system, we use the Weeks-Chandler-Anderson (WCA)~\cite{weeks1971} potential. This takes the form: 
\begin{align}
\label{eqWCA}
u_\mathrm{wca}(r_{ij}) &=
\begin{cases}
4 \varepsilon_\mathrm{wca}[(\frac{\sigma_{ij}}{r})^{12} - (\frac{\sigma}{r})^6] + \varepsilon_\mathrm{wca}  & \ r \le 2^{\frac{1}{6}}\sigma_{ij} \\[2px]
0 &\ r > 2^{\frac{1}{6}}\sigma_{ij}
\end{cases}
\end{align}
where 
$\varepsilon_\mathrm{wca}=10 k_BT$ is the interaction energy and $\sigma_{ij} = (\sigma_i + \sigma_j) / 2$.

We added the dipole-dipole interaction shown in Eq. \ref{eqUDipole}, the Ewald sum for which is implemented with the KSpace package in LAMMPS~\cite{plimpton1995}. Here $\gamma=\gamma_\mathrm{sim}$ controls the strength of  the dipolar interaction. The interaction potential for the simulations then reads

\begin{equation}
u_\mathrm{sim}(r,\theta) = u_\mathrm{wca}(r) + u_\mathrm{dip}(r,\theta).
\label{eqUSimulation}
\end{equation}

Here we quote simulation results in reduced Lennard-Jones units, that is to say the unit of length is the median diameter $\sigma_3$, and and time is in units of $\sqrt{m \sigma_{3}^2/\epsilon_\mathrm{wca}}$ where $m$ is the mass of a particle. We use the Barker-Henderson effective hard sphere diameter of the WCA component of the interaction to determine the effective volume fraction $\phi_\mathrm{eff}$ in order to match the experiments,  ie $\phi_\mathrm{eff} = (\pi / 6V) \Sigma_{i=1,5} n_i \sigma_{i,\mathrm{eff}}^3$ where  $V$ is the volume, $n_i$ is the number of particles and $\sigma_{i,\mathrm{eff}}$ is the effective hard sphere diameter of the $i$th species. Each simulation run includes at least 1000 particles for $\phi=0.1$ and 3000 for $\phi=0.3$.

To prepare the system, the potential energy of random coordinates is minimised under Eq.~\ref{eqUSimulation} to remove overlaps between particles. Then the system is evolved according to the Langevin dynamics. To match the timescales to the experimental system when it is out of equilibrium, we determine the Brownian time in the simulations $\tau_B^\mathrm{sim} = 0.979 \approx 1$. We therefore run the system for 200 time units prior to measurement which matches well to the 20 minutes in the case of the experiments. For equilibrium states ($\gamma<10,12$) for $\phi=0.1$ and 0.3 respectively, we equilibrate the simulations for 100 time units.

\subsection{Bond order parameters for dipolar colloids}
\label{sectionBond}

One method to quantify the angular correlation between particles as a function of external electric field strength is the string fluid order parameter~\cite{li2010}. This has already been shown to be sensitive to variation in field strength by Li \emph{et al.}~\cite{li2010}. This is calculated by finding the angle $\theta$ made by a reference particle with its two nearest neighbors [see Fig. \ref{figBopString}(c) inset]. Here we take $\langle \cos^2\theta \rangle$ as the order parameter for the string fluid. For a system consisting of perfect strings,
 $\langle \cos^2\theta \rangle$ = 1.

\subsection{Two--  and three--body correlation functions}
\label{section23}

We also calculated the two--body spatial correlation function, the radial distribution function $g_2(r)$. 
In addition to the isotropic $g_2(r)$, we consider $g_{2xy}(r)$ which measures correlations in the $xy$ plane perpendicular to the field and $g_{2z}(z)$, which measures correlations in the $z$ direction along the field. For $g_{2xy}(r)$, only pairs of particles that are perpendicular to the $xy$ plane with a tolerance of $\pm 5^\circ$ were used. In the case of $g_{2z}(z)$, particle pairs that are parallel to the $z$-axis were chosen with a $\pm 5^\circ$ tolerance.

We also consider the 3--body spatial correlation function $g_3$. Now this depends on the positions of three particles $1,2,3$, i.e. three vectors, eg $g_3(\mathbf{r}_{12},\mathbf{r}_{23},\mathbf{r}_{31})$ with the numbers reflecting the three particles. Here we elect to simplify our representation to the case where we fix two of the distances (to the particle diameter $\sigma$) so that the 3--body correlation function is plotted as a function of angle between them $g_3(r_{12}=\sigma,{r}_{23}=\sigma,\eta)$ where $\eta$ is the angle between $\mathbf{r}_{12}$ and $ \mathbf{r}_{23}$.

\subsection{Topological Cluster Classification}
\label{sectionTCC}

As discussed in the introduction, the topological cluster classification identifies local geometric motifs whose bond topology (defined here through a modified Voronoi decomposition) is identical to that of minimum energy clusters of a specific size~\cite{malins2013tcc}. These clusters are identified using the energy optimization algorithm GMIN which uses \emph{basin-hopping} to find the local energy minimum that corresponds to a specific configuration for a number of particles in isolation~\cite{wales1997}.  Now such minimum energy clusters require an attractive interaction, and therefore to investigate the effect of the dipolar interaction, clusters were determined for a dipole added to a Lennard--Jones interaction~\cite{skipper2024}. That is to say, the interaction potential for which the minimum energy clusters were determined was

\begin{equation}
u_\mathrm{tcc}(r,\theta) = u_\mathrm{lj}(r) + u_\mathrm{dip}(r,\theta).
\label{eqUTCC}
\end{equation}

\noindent
where $ u_\mathrm{lj}$ is the Lennard--Jones (LJ) interaction. Although the attractive Lennard--Jones contribution is not part of the experimental (or simulated) system we consider here, the importance of packing effects on the structure of liquids and dense fluids has a long history~\cite{barker1976}. The seminal work of Weeks, Chandler and Anderson~\cite{weeks1971} established the role of packing in the structure of dense fluids even in the absence of attractions. It has recently been shown that the higher--order structure of dense hard spheres is closely related to those of attractive systems ~\cite{robinson2019prl,robinson2019pre}. In fact, the higher--order structure of the Lennard--Jones and WCA systems, along with hard spheres, as determined through the topological cluster classification are rather similar~\cite{taffs2010jcp}. We therefore expect that the use of dipolar--LJ clusters will likewise be reasonable here and in any case will provide a suitable measure of the change in the fluid structure under the electric field.

Those resulting clusters which are rigid~\cite{skipper2024} are shown in Fig. \ref{figTCCClusters}. In the case of zero field, we have the minimum energy Lennard--Jones clusters 8B, 9B, 10B, 11C and 12B~\cite{wales,doye1995}. Upon application of the field, the clusters elongate in the $z$-direction. These stretched clusters are denoted S and are based on five--membered rings (9S, 10S, 11S and 12S). Further application of the field leads to clusters which are based on the 6A octahedron. These are denoted O (11O and 12O). Increasing the field still further leads to spiral clusters denoted P (9PAA and 10PAA). Additional  minimum energy rigid clusters of particles interacting according to Eq. \ref{eqUTCC} exist ~\cite{skipper2024}, but here we focus on those we find in our experiments and simulations. In our analysis of the TCC clusters, we set the Voronoi parameter $f_c=0.82$~\cite{malins2013tcc}. Further details are of the identification of the clusters with basin hopping and inplementation in the TCC are available in Ref.~\cite{skipper2024}.

\subsection{Orientation of anisotropic clusters}
\label{sectionOrientation}

Since the dipolar interaction is anisotropic, the clusters found by the TCC may have a preferred orientation with respect to the direction  of the electric field. We consider clusters found in both experiments and simulations. We calculated the principal axis by taking the eigenvector of the inertia tensor of the cluster with the largest eigenvalue. We then determine the angle made by the principal axis  with respect to the direction of the electric field. To quantify this angle distribution, we use an order parameter commonly used for liquid crystals,

\begin{equation}
\langle P_2(\cos\psi) \rangle = \left\langle \frac{3}{2}\cos^2\psi- \frac{1}{2} \right\rangle
\label{eqString}
\end{equation}

\noindent
where $\psi$ is the angle made by the principal axis of each cluster with the direction of the electric field~\cite{selinger2016}. In the isotropic case when the field is switched off, there should be no preferred orientation. In the perfectly aligned state, all the clusters should lie parallel to the electric field so this order parameter is $1$. We note that alignment with the electric field assumes that the clusters are isolated and free to rotate. The system can exhibit a fluid-body centred tetragonal crystal phase coexistence. Some clusters (8O, 11O and 12O) are in fact compatible with this crystal structure and may lie perpendicular to the field, in which case $P_2(\psi=\pi/2) = -1/2$.

\section{Results}
\label{sectionResults}

\begin{figure*}
\begin{center}
\includegraphics[width=125mm]{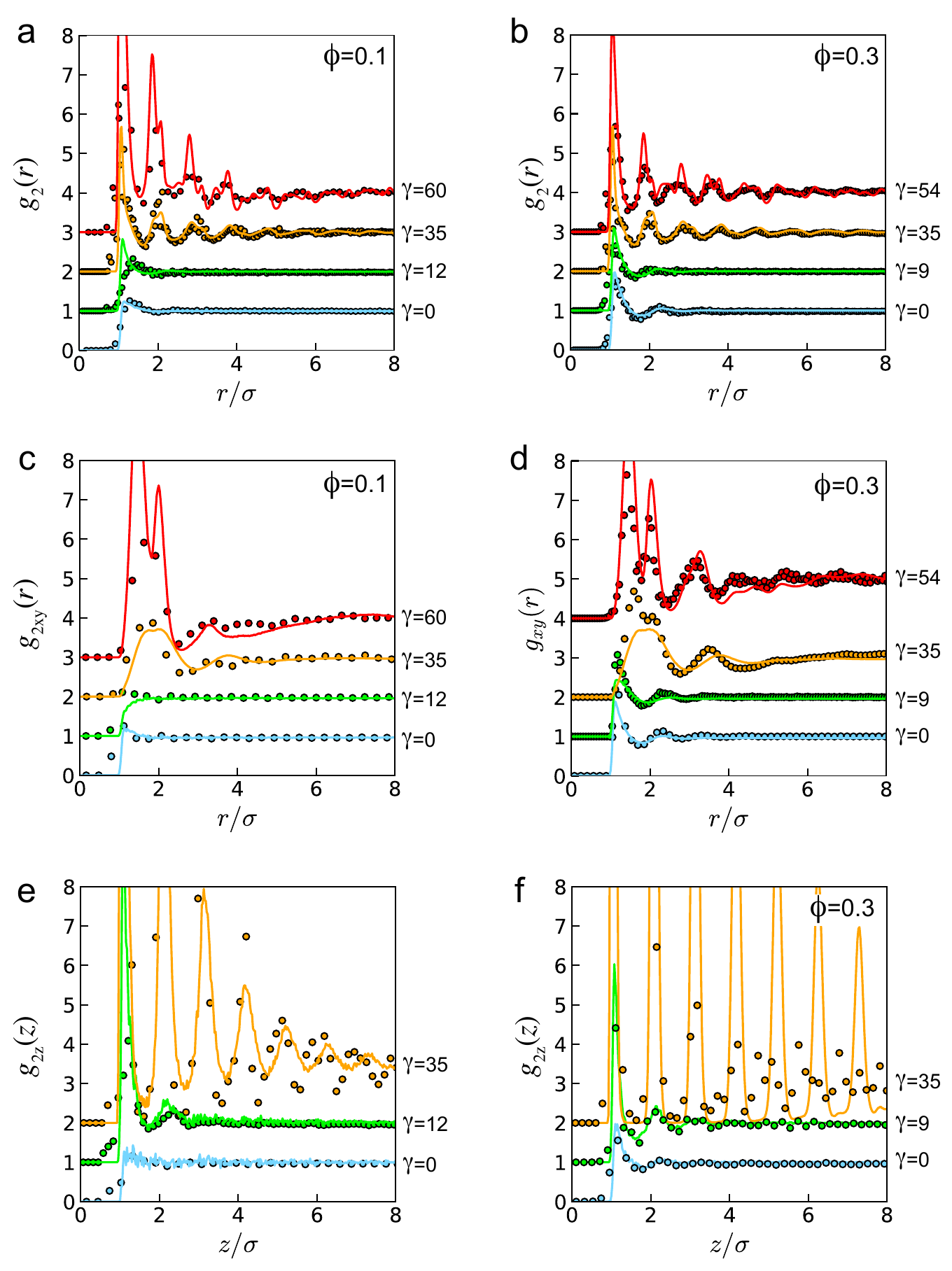}
\end{center}
\caption{Pair correlations  in the colloidal dipolar system. Data are shown for both experiment (points) and computer simulations (lines) at volume fraction (a), (c), (e) $\phi = 0.1$ and (b), (d), (f) $\phi =0.3$. Data are taken for different field strengths expressed through the parameter $\gamma$ as indicated. Data are offset for clarity. (a) and (b) show $g_2(r)$, (c) and (d) show $g_{2xy}(r)$ by considering correlations in the $xy$ plane, (e) and (f) show $g_{2z}(r)$ 
where correlations are taken along the $z$ axis.}
\label{figRadial}
\end{figure*}

We now present both experimental and simulation results for colloidal dipolar fluids at volume fractions $\phi=0.1$ and $\phi=0.3$. We investigate the bond order parameters, two and three body correlation functions, and populations of minimum energy clusters and the orientations of these clusters. These quantities are considered as a function of dipole strength $\gamma$. We convert external field strength to the dipole strength $\gamma$ using Eq. \ref{eqGamma}, which we use across experiments (Eq. \ref{eqDipoleMoment}), simulations (Eq. \ref{eqUSimulation}) and minimum energy clusters (Eq. \ref{eqUTCC}).

\subsection{Bond order parameter analysis of the string fluid}
\label{sectionBopAnalysis}

In our study, with the string fluid order parameter $\langle \cos^2 \theta \rangle$ (see Sec. \ref{sectionBond}), we explore slightly higher colloid volumes fraction (0.1 and 0.3) than some previous work~\cite{li2010}. Our results are shown in Fig. \ref{figBopString} and for the string order parameter, we find good  agreement between experiment and simulation  if we scale the values of $\gamma$ in the experiment by 1.2, which we do henceforth. The value of  $\langle \cos^2 \theta \rangle$ is small for both experiment and simulation in the case of zero field, and then shows a significant increase with field strength to a value of  $\langle \cos^2{\theta}\rangle \sim 0.8$ for $\phi=0.1$ and $\sim 0.65$ for $\phi=0.3$.

Upon increasing the field strength such that $\gamma\approx12$ and $\approx10$ for volume fraction $\phi=0.1$ and $0.3$ respectively, the string fluid becomes metastable to fluid-body-centered tetragonal phase coexistence~\cite{hynninen2005}. Since the system is now out of equilibrium, it is possible that small structural differences between the experiments and simulations may emerge, such as the small increase in $\langle \cos^2{\theta}\rangle$ in the experiments with respect to the simulations for  $\phi=0.1$.

\subsection{Pair Correlation Function}
\label{sectionPair}

We continue our analysis by considering pair correlations in the form of the radial distribution function $g_2(r)$ in Fig.~\ref{figRadial}. The orientationally averaged $g_2(r)$ is shown in Figs.~\ref{figRadial}(a) and (b) for volume fraction $\phi=0.1$ and 0.3 respectively for both experiment (data points) and simulation (lines). At zero field strength, in the hard sphere limit, we see reasonable agreement between experiment and simulation (as has been noted previously)~\cite{royall2023}. The slightly higher first peaks of the simulation data may be attributed to the particle tracking errors in the experiments~\cite{royall2007jcp}. For weak field strengths ($\gamma=9$), again we see comparable agreement between experiment and simulation to that of the hard sphere case.

As above in Sec.~\ref{sectionBopAnalysis}, at high field strengths when the system falls out of equilibrium, we see some discrepancy between experiment and simulation. In particular, we see stronger peaks, ie more ordering, in the simulation data for $\gamma=12$ and $19$ for $\phi=0.1$. We believe that this difference is too large to be attributed to tracking error. % and polydispersity. 
Interestingly, the difference between experiment and simulation is rather less significant in the case of $\phi=0.3$. We return to consider possible influences in the structure out of equilibrium in Sec. \ref{sectionDiscussion} below.

Since our system is anisotropic due to the external field, we expect to see some corresponding differences in structure between the plane perpendicular $xy$ and the direction parallel $z$ to the field. To explore this we now consider the pair correlation function in the $xy$ plane $g_{2xy}(r)$ and $z$ direction $g_{2z}(z)$. We expect that the formation of the string fluid may lead to significant structuring in the field direction, while in the perpendicular $xy$ plane, there is repulsion between particles exactly in plane, but out--of--plane attractions lead to aggregation of the strings~\cite{messina2023} and ultimately the formation of the bct crystal. In Fig.~\ref{figRadial}(c,d), we show the pair correlations in the perpendicular plane for volume fraction $\phi=0.1$ and $0.3$ respectively. In the case of $\phi=0.1$ and $0.3$ at zero field strength we see good agreement between experiment and simulation. For a weak field at low volume fraction, there is some evidence of enhanced repulsion in the experiments with respect to the simulations, but this is not seen at higher volume fraction. Upon increasing the field strength such that $\gamma=35$, we see a reasonable agreement between experiment and simulation. This is echoed in the higher volume fraction case ($\gamma=35$) with again stronger peaks in the experiment. At the highest field strength, $\gamma=60$ and $54$ for $\phi=0.1$ and $0.3$ respectively, the agreement between experiment and simulation is rather good.

Turning to the case of the correlations along the field direction, $g_{2z}(z)$  [Figs.~\ref{figRadial}(e) and (f)], for zero field strength, we recall that we may expect to see evidence of significant ordering upon application of the electric field, as the string fluid develops. At zero and low field strength, correlations are comparatively weak for both $\phi=0.1$ and $0.3$ [though note the scale on the $y$-axis in Figs.~\ref{figRadial}(e) and (f)] and experiment and simulation are well matched. At higher field strength we find significantly stronger peaks in the case of the simulations for both volume fractions. This may be due to tracking errors, where we note the the resolution of the confocal microscope is reduced in the $z$ direction and thus tracking errors may be more severe here.

Summarising the behaviour we have uncovered through our analysis of the pair correlations, we find overall reasonably good agreement between experiment and simulation 
We suggest that discrepancies between experiment and simulation may be attributed to particle tracking errors (particularly in the case of $g_{2z}(z)$ at higher field strength) ~\cite{royall2007jcp,royall2023}.

\subsection{Three-body Correlations}
\label{sectionThree}

\begin{figure*}
\begin{center}
\includegraphics[width=150mm]{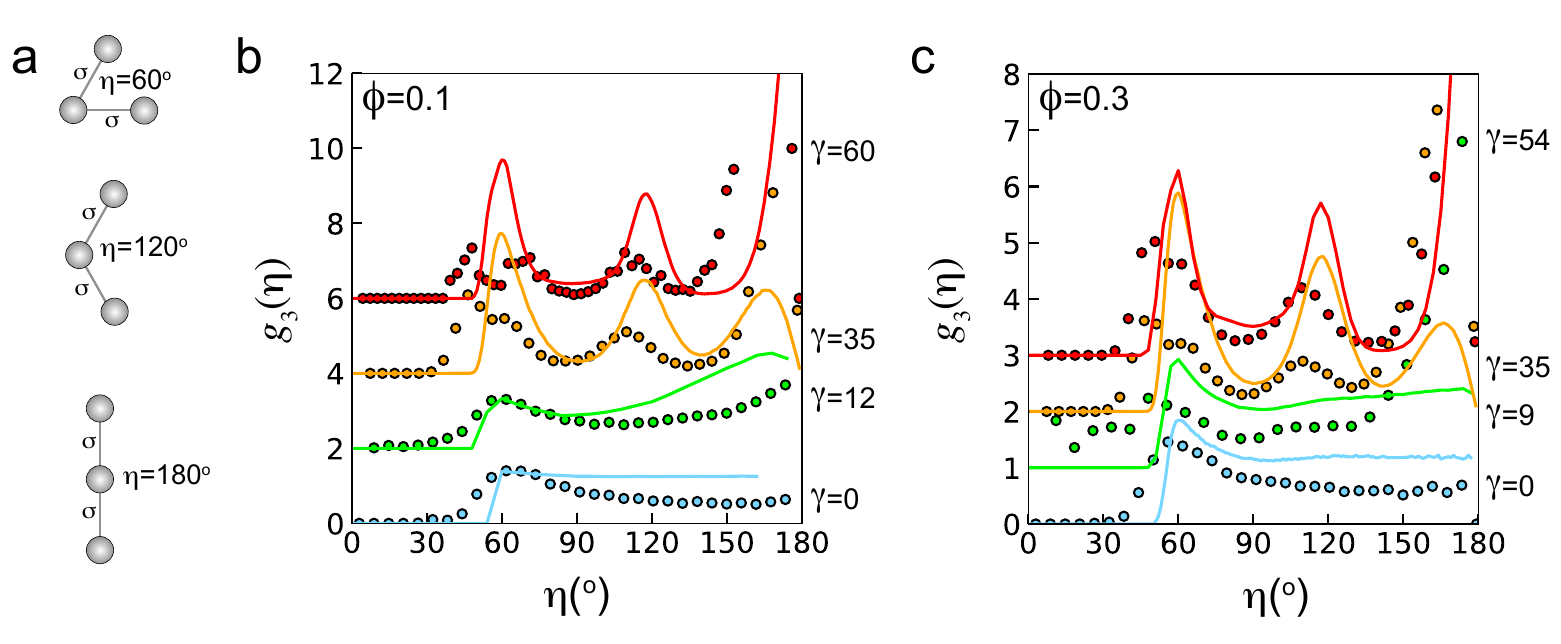}
\end{center}
\caption{The three--body correlation function $g_3(\eta)$. 
(a) Schematic indications of geometries of interest, with values of the bond angle $\eta=60^{\circ}$.120$^{\circ}$ and $180^{\circ}$
(b) $g_3(\eta)$ is plotted for volume fraction $\phi=0.1$.
(c) $g_3(\eta)$ for volume fraction $\phi=0.1$.
In (b) and (c), data points are experimental data and lines are simulation data.}
\label{figG3}
\end{figure*}

As noted above, particle--resolved studies lends itself to analysis of higher--order correlations~\cite{vanblaaderen1995,royall2008,leocmach2013,royall2015physrep,royall2023} and here we begin with the three--body correlation function $g_3$. This may be represented in a variety of ways and here we chose to show the dependence of $g_3$ upon the angle $\eta$ between two particles with respect to a particle of interest as shown in Fig.~\ref{figG3}(a). We set the distance between the two particles and the particle of interest to be the diameter, so that the quantity plotted is $g_3(\eta)$.

For both $\phi=0.1$ and $0.3$, at zero and low field strength, a large and broad peak appears at 
$\eta\approx60^{\circ}$ which is consistent with an isotropic system where the interaction is angle independent~\cite{coslovich2013jcp}. This is more prevalent in our experiments than in the simulations. Upon applying the field, a peak at $\eta\approx180^{\circ}$ emerges, relating presumably to the development of the string fluid. For $\phi=0.1, \gamma=12$, This is more prevalent in the simulations than in the experiments, whereas at the slightly weaker field  $\gamma=9$ for $\phi=0.3$, the experiments exhibit a higher peak. When the system falls out of equilibrium, for both volume fractions, for we see rather enhanced ordering in the simulations for $\gamma=35$ with a much stronger peak around $\eta\approx120^{\circ}$ . This may be related to a more rapid demixing towards a bct crystal. The peak towards  $\eta\approx180^{\circ}$ is in fact weaker in the simulations than the experiments. For $\phi=0.1$, this is consistent with a slightly higher development of string-like structure in the experiments. At the highest field strength, both experiments and simulations show an increased degree of ordering, as evidenced in the peaks at $\eta\approx60^{\circ}$, $120^{\circ}$ and $180^{\circ}$. Again, the peaks seem somewhat more pronounced in the simulations.

\subsection{Populations of minimum energy clusters}
\label{sectionTCCAnalysis}

\begin{figure*}
\begin{center}
\includegraphics[width=130mm]{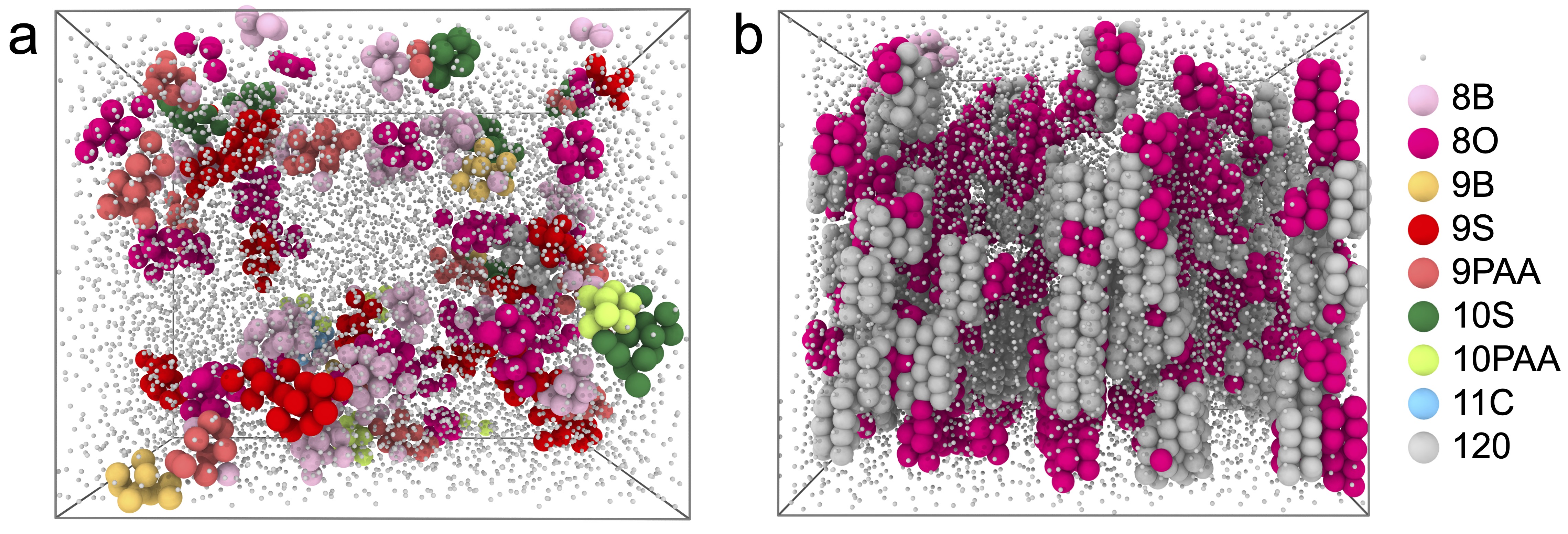}
\end{center}
\caption{
Rendering of TCC clusters using experimental coordinate data. 
Key indicates the particular cluster a particle is in. Particles are rendered in the largest cluster in which they are identified.
Small grey particles are not identified in any of the clusters considered.
(a) $\phi=0.3,\gamma=0$.
(b) $\phi=0.3,\gamma=54$.
}
\label{figPretty}
\end{figure*}

\begin{figure*}
\begin{center}
\includegraphics[width=125mm]{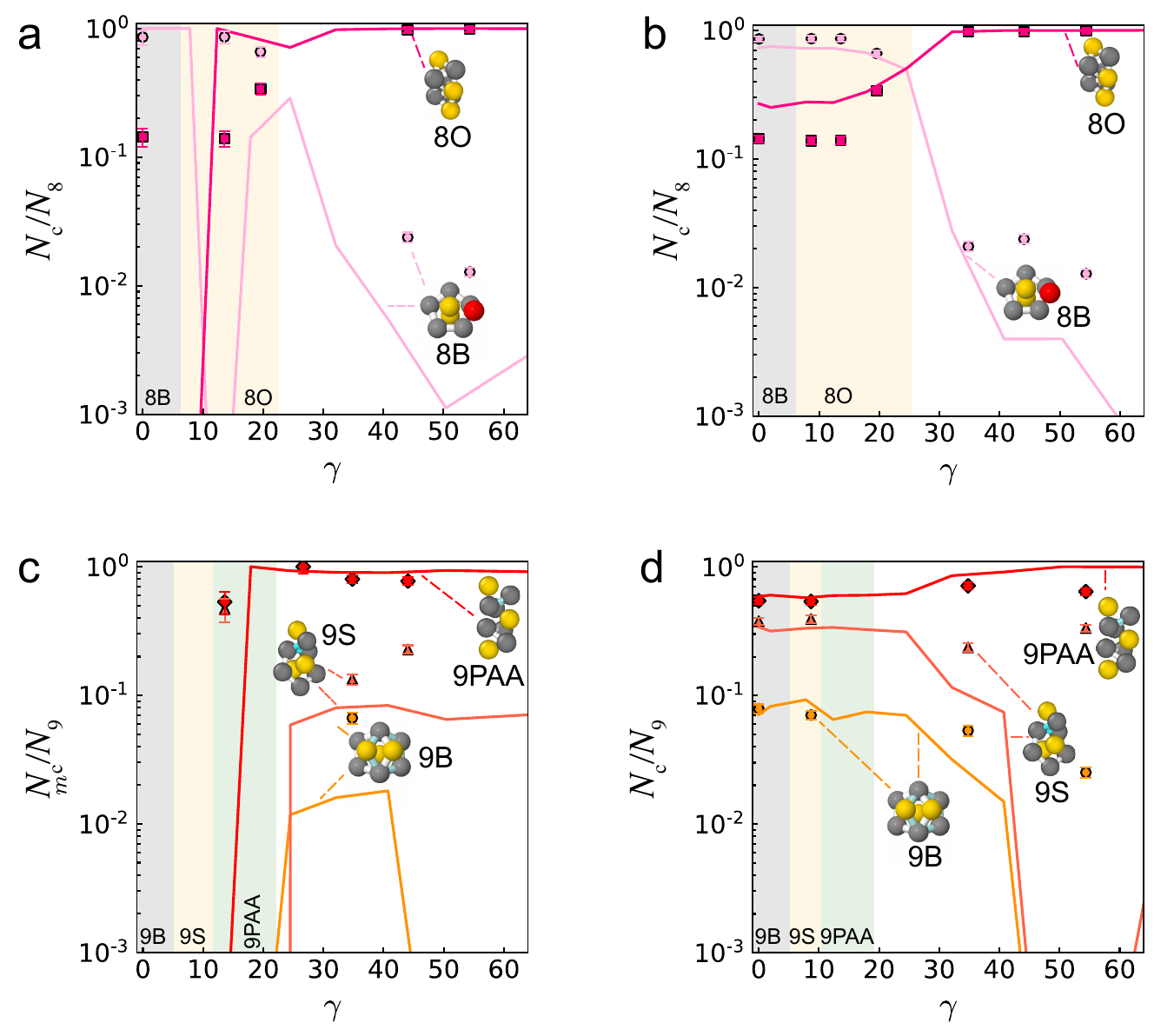}
\end{center}
\caption{Populations of smaller minimum energy clusters detected by the topological cluster classification as a function of dipole strength. The number of particles detected in a given cluster $N_c$ is scaled by the number of particles in a cluster of that size, $N_m$. Data points denote experiment and lines are computer simulation. Shading denotes the change in cluster topology which minimizes the energy at different values of the dipole strength $\gamma$ as indicated~\cite{skipper2024}. White regions of graphs denote values of $\gamma$ where the minimum energy clusters are not rigid, that is to say, single-particle or two-particle wide strings are formed, or other non-rigid structures~\cite{skipper2024}.
Data are shown for different cluster sizes $m$ and volume fractions as follows. 
(a) Cluster size $m=8$, volume fraction $\phi=0.1$.
(b) $m=8$,  $\phi=0.3$.
(c) $m=9$,  $\phi=0.1$.
(d) $m=9$,  $\phi=0.3$.
}
\label{figTCCPopulationA}
\end{figure*}

We now move to still higher--order spatial correlations and consider the dipolar--Lennard--Jones minimum energy clusters identified by the topological cluster classification (Fig.~\ref{figTCCClusters})~\cite{skipper2024}. In Fig.~\ref{figPretty}, to give an overview of the response of the cluster population to the electric field, we render two example state points. These are $\phi=0.3,\gamma=0$ in Fig.~\ref{figPretty}(a) and  $\phi=0.3,\gamma=54$. We see some Lennard-Jones clusters in the case of zero field (8B, 9B and 11C). For the higher field strength, we see an organisation into columns (related to the gradual demixing towards the bct crystal). These are dominated by 8O and 12O clusters. We now proceed to analyse the cluster population in our system in more detail.

\begin{figure*}
\begin{center}
\includegraphics[width=165mm]{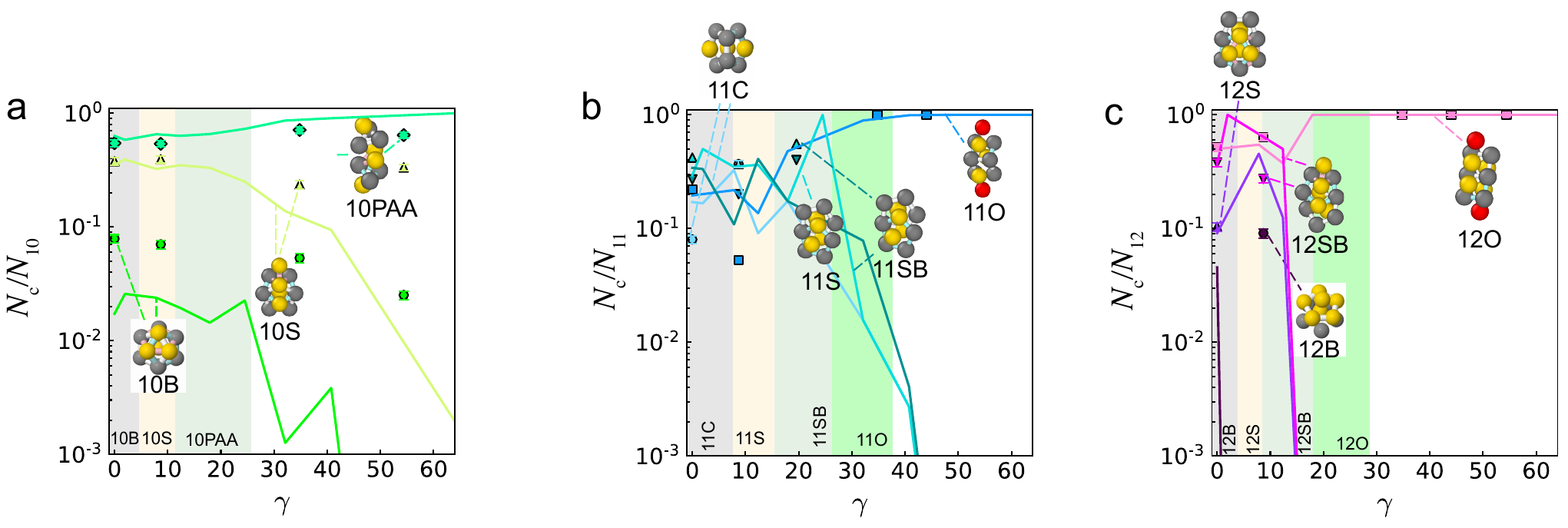}
\end{center}
\caption{Populations of larger minimum energy clusters detected by the topological cluster classification as a function of dipole strength. As in Fig. \ref{figTCCPopulationB}, the number of particles detected in a given cluster $N_c$ is scaled by the number of particles in a cluster of that size, $N_m$. Data points denote experiment and lines are computer simulation. Shading denotes the change in cluster topology which minimizes the energy at different values of the dipole strength $\gamma$ as indicated~\cite{skipper2024}. White regions of graphs denote values of $\gamma$ where the minimum energy clusters are not rigid. Data are shown for different cluster sizes $m$. 
(a) Cluster size $m=11$, 
(b) $m=12$. 
Here volume fraction $\phi=0.3$.}
\label{figTCCPopulationB}
\end{figure*}

To facilitate comparison between experiment and simulation, in each figure we fix the number of particles in the cluster in question. Now the topology of the minimum energy cluster changes upon increasing the dipolar contribution~\cite{skipper2024}. In Fig.~\ref{figTCCPopulationA}, we consider 8 and 9--membered clusters for volume fraction $\phi=0.1$ and $0.3$.  For larger clusters, there are fewer statistics at lower volume fraction and therefore, we focus on $\phi=0.3$ in Fig.~\ref{figTCCPopulationB}. 
In line with the discussion in Sec.~\ref{sectionTCC}, and previous work~\cite{robinson2019prl,taffs2013,taffs2010jpcm}, despite the lack of attraction between the hard spheres in the case for zero field strength, we nevertheless expect to find some clusters due to packing effects.

Naively, one might expect that zero and low field strength would correspond to minimum energy clusters for the Lennard--Jones interaction as shown in Fig. \ref{figTCCClusters} and known from studies with hard spheres~\cite{taffs2013,royall2018jcp,royall2023,robinson2019pre}, and that increasing the field strength might lead to a cascade of clusters of increasing elongation as indicated in Fig. \ref{figTCCClusters}(b). This turns out to be the case.

For the smallest size of cluster we consider, $m=8$, indeed we see this trend with the Lennard--Jones minimum energy cluster 8B giving way to the 8O which minimises the potential energy for the dipolar system for both volume fractions [Fig.~\ref{figTCCPopulationA}(a,b)]. Notably, experiment and simulation appear reasonably well--matched in the lower field case that the system is in equilibrium ($\gamma \leq 12, \phi=0.1$ and $\gamma \leq 10, \phi=0.3$). By this we mean that discrepancies are well within an order of magnitude (note that, for an 8-membered cluster to be successfully identified, failure to identify only one of the 8 particles will lead to the cluster not being identified, so ``agreement'' between experiment here is inevitably rather less stringent than in the case of pair correlations, say). The more significant discrepancy between the experiments and simulations emerges. At higher field strength, $\gamma>30$ in both volume fractions $\phi=0.1$ and $0.3$, we see an increase in the 8B modified pentagonal bipyramid population in the experiments compared to the simulations. Now this structure exhibits fivefold symmetry, and as such, is associated with non--crystalline ordering~\cite{frank1952}. We have noted above that, in the non-equilibrium conditions at higher field strength, the simulations appear more ordered. If the ordering in the simulations is crystalline, then a lack of five-fold symmetry with respect to the experiments would seem to be reasonable.

In the case of 9-particle clusters, the situation is more complex, as we consider three clusters, 9B, 9S and 9PAA.  We find it instructive to start our analysis with $\phi=0.3$ Fig.~\ref{figTCCPopulationA}(d)]. At low field strength, $\gamma=0$ and $9$, there is a rather high population of 9S and 9PAA in both experiment and simulation. This dominates over the minimum energy cluster for the Lennard-Jones system, 9B which has two five--membered rings~\cite{malins2013tcc}. At higher field strengths, the population of 9B drops rather precipitously in the simulations, but less so in the experiments. This is consistent with the case of $m=8$ above, where the 8B cluster which, like the 9B, has a degree a fivefold symmetry is preferred. Notably though, the 9PAA dominates at all field strengths, which is not expected from energy considerations as it is the minimum energy cluster only for $12  \lesssim \gamma \lesssim 21$. Now the 9PAA cluster is polytetrahedral in structure, and is enlongated with respect to the 9B. This is consistent with it being intermediate between the compact 9B in the case of zero field and strings of particles in the case of a strong field. From a structural perspective the 9S \emph{stretched} polytetrahedron is intermediate between 9B and 9PAA in that it is the minimum energy structure for $7  \lesssim \gamma \lesssim 12$. Its population is also intermediate between 9B and 9PAA for both experiments and simulations. For $\phi=0.1$ [Fig.~\ref{figTCCPopulationA}(c)], there are rather fewer clusters identified, and none at all at low field strength. It is often the case that there are fewer (larger) clusters at lower volume fraction~\cite{malins2009,malins2010,klix2013,royall2023}, so this in itself is reasonable. Like the case of $\phi=0.3$, the 9PAA dominates at all field strengths. At high field strength, we see more of the clusters with five--membered rings, 9B and 9S in the experiments than in the simulations.

We now consider larger clusters in Fig. \ref{figTCCPopulationB}(a,b,c). The statistics for larger clusters at the lower volume fraction are rather poor and therefore here we focus on a volume fraction  $\phi=0.3$ respectively. Turning to $10$-membered clusters [Fig.~\ref{figTCCPopulationB}(a)], we see a somewhat similar behaviour to the $m=9$ case. For both experiments and simulations, we see some of the Lennard--Jones minimum energy cluster, the defective icosahedron 10B at low field strength as one might expect. Like the 9-membered clusters, the 10B is present only in small quantities with the 10PAA dominant at all field strengths and 10S intermediate. In experiments the population of 10B (which exhibits some fivefold symmetry) is higher in the experiments than in the simulations, echoing the case for 8B in Figs.~\ref{figTCCPopulationA}(a) and (b). Meanwhile the populations of 10S and 10PAA are comparable between experiments and simulations.

For 11-membered clusters, at low field strength, in experiment, we find 11S, 11SB and 11O dominating with rather less 11C, the latter being the minimum energy Lennard--Jones cluster, like the smaller Lennard-Jones clusters the 11C has a number of five-membered rings which may underly its small population. At higher field strength 11O dominates, similar to the case for 8O in $m=8$. However for $m=11$ the additional two structures 11S and 11SB seem to lie between 11C and 11O in population, and their population falls away at higher field strength. This is broadly consistent with expectations of 11O dominating at higher field strength (it is the minimum energy cluster for $27 < \gamma < 38$.

Finally, the 12-membered clusters are shown in Fig. \ref{figTCCPopulationB}(b). Here in the experimental data, 12O, which is the minimum energy cluster for $19 < \gamma < 31$ is the most popular cluster at all field strengths. We see some 12B and 12SB at weak field strengths, but these vanish for field strengths greater than $\gamma = 15$.  The simulations exhibit the same qualitative trend, of 12O dominating.

\subsection{Cluster Orientation}
\label{sectionOrientationResults}

We have observed that even at zero dipole strength, some of the minimum energy dipolar clusters (such as 10PAA at $\phi =0.3$) are present in our system [Fig.~\ref{figTCCPopulationA}(f)]. Now the dipolar--Lennard--Jones clusters that we consider (Fig.~\ref{figTCCClusters}) are aligned with the dipolar interactions, and thus with the electric field. It is therefore reasonable to suppose that the clusters might exhibit some alignment with the field, and that this would increase as a function of field strength.

\begin{figure*}
\begin{center}
\includegraphics[width=130mm]{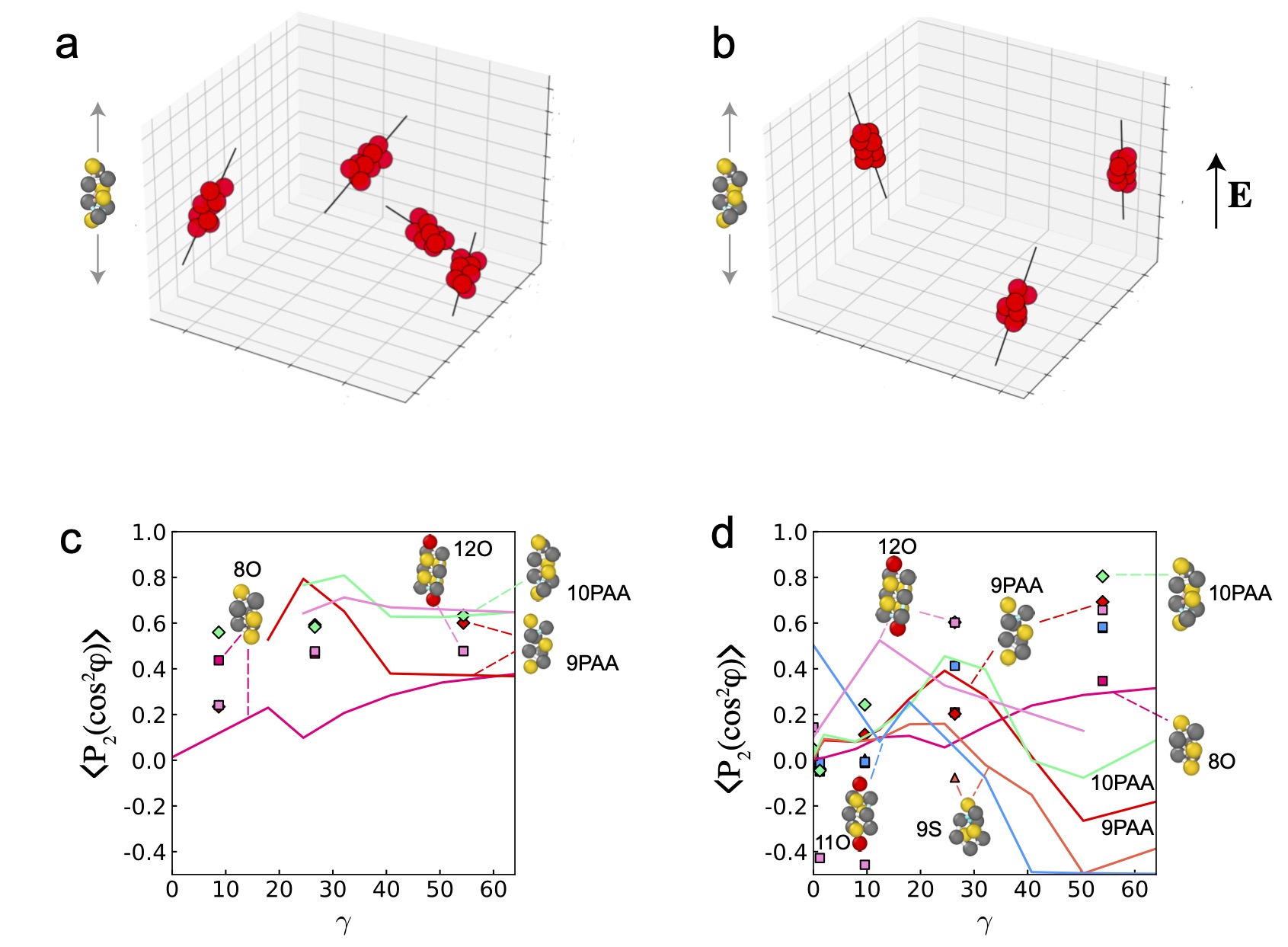}
\end{center}
\caption{
Orientation of anisotropic clusters with the electric field.
(a) shows a 3d plot of the clusters (10PAA) found in the experiments at $\phi=0.3$ at zero field, where the principal axis (indicated as the black line) of each cluster does not align with the electric field. Whereas in (b) these 10PAA clusters showed a higher degree of alignment along the field ($z$-axis) as the field takes a value of $\gamma=40$.
(c,d) $\langle P_2(\cos\psi) \rangle $ of the principal axis is plotted as a function of reduced field strength, for the anisotropic clusters 8O, 9PAA, 9S , 10PAA, 11O and 12O. for the two volume fractions under consideration.
(c) $\phi =0.1$ and (d) $\phi =0.3$. Data points are experimental data and lines are computer simulation.
Shading denotes the system falling out of equilibrium~\cite{hynninen2005}.
}
\label{figOrientation}
\end{figure*}

We therefore probe the orientation of some dipolar--Lennard--Jones clusters using our method described above in Sec.~\ref{sectionOrientation}. 
The principle is illustrated in Fig.~\ref{figOrientation}(a) and (b). Here, renderings of experimental data where clusters identified as 10PAA for volume fraction $\phi=0.3$ are shown. In the case of zero field strength [Fig.~\ref{figOrientation}(a)], no preference in cluster orientation is seen. For $\gamma=54$ [Fig.~\ref{figOrientation}(b)], we see that the clusters have oriented with the field. We now explore this phenomenon quantitatively. In Fig.~\ref{figOrientation}(c) and (d), for volume fractions $\phi=0.1$ and $0.3$ we plot the degree of alignment with the field for experimental data (data points) and simulation (lines).

We begin our analysis with $\phi=0.1$ [Fig.~\ref{figOrientation}(c)]. As exemplified by 8O (for which we have the best statistics), there is a trend towards more alignment with the field as its strength is increased. The experimental data in general shows an increase in alignment, while for the simulations, 9PAA, 10PAA and 12O show some non-monotonicity in their response to the field. Turning to the case of higher volume fraction, in Fig.~\ref{figOrientation}(d), we see a similar behaviour. Experimental data all show an increase in alignment with the field. Contrasting with this, in the simulations, 11O even seems to approach to $\langle P_2(\cos \psi) \rangle = -1/2$, the case for alignment in the $xy$ plane. This may be due to the cluster being identified in the bct crystal in which it may lie perpendicular to the field.

\section{Time-Evolution}
\label{sectionAging}

\begin{figure*}
\begin{center}
\includegraphics[width=130mm]{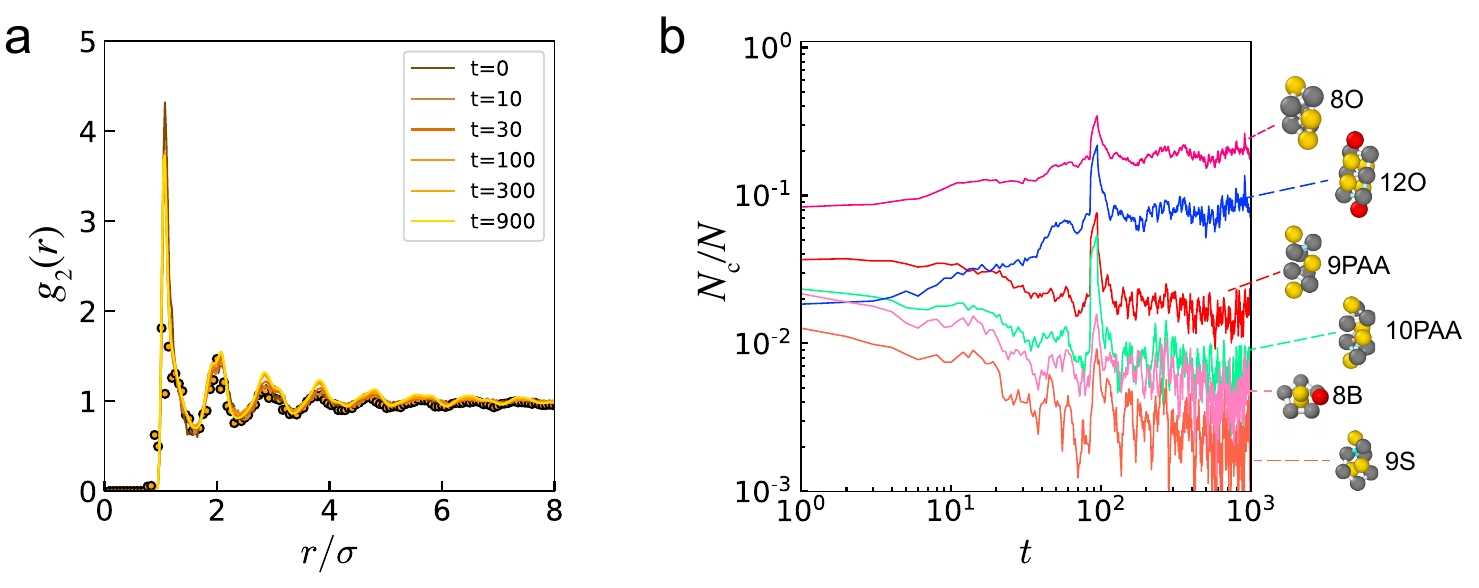}
\end{center}
\caption{Aging in the phase separation regime.
Here we consider the state point $\phi=0.3, \gamma=32$.
(a) Radial distribution functions $g_2(r)$ are plotted for the times indicated. These are
compared with experimental data (for a waiting time around 200 $\tau_B$). Experimental data [plotted in Fig.~\ref{figRadial}(a)] and simulations data taken for $t=0,30,100,300,900$ time units.
(b) Time-evolution of selected TCC clusters for the simulation. Here time is in Lennard-Jones time units.
}
\label{figAging}
\end{figure*}

We have remarked that in the regime that $\gamma>10,12$ for $\phi=0.1,0.3$, the system becomes metastable to a phase separated fluid-bct crystal state~\cite{hynninen2005}. 
As an outlook, we briefly consider the consequences of time-evolution for our analysis. To this end, we run a longer simulation of 1000 time units that we analysed this from the beginning (without applying any equilibration time after the initial energy minimisation). We consider the $\phi=0.3, \gamma=32$ state point. In Fig.~\ref{figAging}(a), we plot the radial distribution function, and also reproduce the experimental data for this state point from Fig.~\ref{figRadial}(b). The time-evolution of the $g_2(r,t)$  does show some changes, notably a split second peak can be detected longer times. This is compatible with increased ordering~\cite{truskett1998}. The changes in the pair-correlations are, however, rather small.

We now turn to the higher-order correlations and in particular plot the time-evolution of certain TCC clusters as a function of time in Fig.~\ref{figAging}(b). These show rather clear trends following their geometry. We have remarked that the clusters incorporating octahedra, in particular 8O and 12O are compatible with the bct crystal and these are indeed found at high field strength [Fig.~\ref{figPretty}(b)]. The population of these clusters increases with time here which is consistent with the system moving towards the fluid-bct state. Conversely the Lennard-Jones cluster 8B and 9S which also exhibits some five-fold symmetry find their population falling, presumably their symmetry is incompatible with the fluid-bct state. Finally, clusters which are ground states at intermediate field strength, 9PAA and 10PAA~\cite{skipper2024} exhibit relatively little change in their population. This suggests that the TCC maybe a suitable means to probe the time-evolution of the dipolar colloids. A more extensive investigation we leave for the future.

\section{Discussion}
\label{sectionDiscussion}

We now discuss our findings figure by figure.

(\emph{i})
Figure \ref{figBopString} shows confocal microscopy images of our system at volume fraction $\phi$=0.1, taken along the $xy$ plane (perpendicular to the direction of the electric field) and $yz$ plane (along the direction of field) for both zero dipole strength and at maximum dipole strength at $\gamma$=45. We see string formation along the direction of the field at $\gamma$=45.

Now our system becomes metastable to fluid-bct crystal phase coexistence at $\gamma \geq 10$ and $12$ for volume fraction $\phi=0.1$ and $\phi=0.3$ respectively. Under these conditions, given that throughout most of this work, we do not treat the time--evolution of the system discrepancies between the experiments and simulations are possible due to each taking a different route through the energy landscape. However, our analysis in Fig.~\ref{figAging} suggests that for two-point correlation functions, effect of the time-evolution is quite weak. For the TCC clusters, we do see some changes, and we discuss this in more detail below.

Figure \ref{figBopString}(c) shows a plot of bond order parameters of the string fluid from both our experiment and simulation results, similar to the study published by Li \emph{et al.}~\cite{li2010}. Here, our simulations (line) and experiments (data points) show good agreement. As the field strength increases, the bond angle $\theta$  tends towards 180$^{\circ}$. As indicated in Fig. \ref{figBopString}(c), the degree of string formation increases which is consistent with the string formation is continuous rather than a sharp transition.

(\emph{ii}) 
In Fig. \ref{figRadial} we plot $g_2(r)$. This is generally in reasonable agreement between computer simulations and experiments for both $\phi$=0.1 and $\phi$=0.3. We can therefore be fairly confident that the simulation model used in our work is a reasonable reflection of our experimental system. We see the emergence of long range order as field strength is increased. This is expected since the confocal images show that as the fluid becomes more structured as the colloids aligned along the field when it is switched on. The pair correlation function $g_2$ does however show significant discrepancies emerge at high field strength. The increased ordering in the simulations may be due to the system being further down the path to forming a bct crystal than is the case in experiment.

The results of $g_{2xy}(r)$ and $g_{2z}(z)$ are both consistent with that of $g_2(r)$ where the height of the first peak increases as field strength is switched on. The significant differences in peak positions and shapes between $g_{2xy}(r)$ and $g_{2z}(z)$ show the fluid structure across the $xy$-plane differs from fluid structure along the $z$-axis. We also observe peak splitting ($g_{2xy}(r)$) from a broad peak into two distinct peaks as field strength increases from $\gamma$=35 to $\gamma$=60/54, showing increase in ordering of fluids across the $xy$-plane. Whereas ordering in the fluid along the $z$-axis occurs above  $\gamma$=12/9. However, fluid structures still show little difference at low field strength, except that the first peak in particular is stronger in the case of the simulations, which we attribute to particle tracking errors. At higher field strengths as the system falls out of equilibrium and starts to order, larger discrepancies emerge, notably in the higher first peak in $g_{2xy}(r)$ and much higher peaks in $g_z(z)$. The latter we attribute to tracking errors.

(\emph{iii}) 
We now consider three--body correlations in Fig. \ref{figG3}. As the dipolar strength increases, the peak at 180$^{\circ}$ increases as expected since more dipolar colloids form strings. However, we also observe peaks at 60$^{\circ}$ and 120$^{\circ}$ increasing with respect to field strength. The triplet correlations in the simulations show increasing structure at higher field strength with respect to the experiments.

(\emph{iv}) 
The renderings in Fig.~\ref{figPretty} and the plots in Figs. \ref{figTCCPopulationA} and \ref{figTCCPopulationB} show the population of dipolar clusters of different geometries and sizes analysed with the TCC. Overall, we find that the clusters we observe in our system follow reasonably those of the dipolar--Lennard--Jones clusters (Fig.~\ref{figTCCClusters}). That is to say, we see more elongated clusters at higher field strengths. In all cases, at high field strength it is the LJ--dipolar cluster that corresponds to the highest dipolar interaction that we find in both experiment and simulation. We find this to be a significant outcome of this work, providing strong evidence in support of modelling colloids in an AC electric field with dipolar interactions.

Minimum energy dipolar clusters by definition imply zero temperature, and are determined by the interaction energy. However, at finite temperature, entropy plays a role. The fact that the cluster population trends largely follow the minimum energy clusters indicates that the behavior of dipolar colloids is is strongly influenced by energetics. This is in stark -- and surprising -- contrast to earlier work which suggested that energy plays only a very limited role in observed cluster populations~\cite{taffs2010jcp}. That work investigated Frank's well--known conjecture that icosahedra ``will be a very common grouping in liquids''~\cite{frank1952}. In fact, at the triple point of the Lennard--Jones system, only one particle in 1000 was found to be in an icosahedron and other 13--membered clusters dominate~\cite{taffs2010jcp}, quite unlike the findings here in which the minimum energy structure dominates at high field strengths. Presumably the strength of the dipolar interactions (which are much larger than eg interactions in the Lennard--Jones system when it is in the liquid state~\cite{taffs2010jcp}) is important here. At high field strength, for $m=8,9,10$ clusters we find more clusters with five--membered rings in the experiments than in the simulations. This we attribute to geometric frustration as the system falls out of equilibrium, with experimental and simulated realisations of the system taking different paths in the energy landscape (see below).

(\emph{v}) Figure \ref{figOrientation} shows that some anisotropic dipolar clusters tend to align along the $z$-axis (parallel with direction of the field) while others in fact orient perpendicular to the field. This is a surprising result and we attribute it to incipient crystallisation in the system. Inspection of Fig.~\ref{figPretty} shows ordering and a structure dominated by 8O and 12O. While in the experiments the 8O and 12O rich regions are oriented vertically, it is possible that in simulation a somewhat different path through the energy landscape is followed.

(\emph{vi}) Figure \ref{figAging} provides an indication of how the time-evolution of the system may be monitored. We see that changes in the radial distribution function are limited, but the time-evolution of the TCC clusters is very much more marked. This suggests that this sort of analysis may be useful to probe kinetic pathways in the dipolar colloidal system.

We now wish to discuss the relevance of our results with higher order structure. Higher-order structure is sensitive to changes in structure in a way the two-point correlation function are not. For example, at low volume fraction, even relatively weak field strengths ($\gamma=12$) cause a dramatic change in the population of, for example the 8O cluster [Fig.~\ref{figTCCPopulationA}(a)] while $g_2(r)$ exhibits rather little change. We can therefore conclude that higher order analysis such as that presented here is much more sensitive to small variation in structure and interactions. Comparing our study with other previously published work which used a TCC analysis of gels and glasses~\cite{royall2008,royall2015jnonxtalsol,leoni2023,jenkinson2017,royall2015jnonxtalsol,royall2012,thijssen2023}, we can conclude that such higher order structural analysis is better at capturing the onset of structural changes in amorphous systems than pair correlations $g_2(r)$, as may be inferred from other work~\cite{vanblaaderen1995,coslovich2011,leocmach2012,leocmach2013}. What is new here is that we have considered a system with anisotropic interactions.

The more significant %major 
discrepancies that we find are in the regime in which the system departs from equilibrium. That is to say, at high field strengths, the system becomes metastable to fluid-bct crystal phase coexistence. Now the early stages of this transition have been investigated recently~\cite{messina2023}. However, in related phenomena, such as the condensation of colloids with an effective attraction to form a gel network, the role of hydrodynamic interactions was found to be very important. Hydrodynamics control not only the timescale for the condensation~\cite{furukawa2010}, but also the higher--order structure of the resulting non-equilibrium gel network~\cite{royall2015prl,degraaf2019}. In (non--equilibrium) gelation of particles with spherically symmetric attractions, experiments exhibit many fewer clusters with fivefold symmetry than do Brownian dynamics computer simulations~\cite{royall2015prl}, quite the opposite trend of what is observed here in Figs.~\ref{figTCCPopulationA} and Figs.~\ref{figTCCPopulationB}. We suggest that careful study, using simulations with hydrodynamic interactions and time--resolved experimental observation along the lines of Ref.~\cite{royall2015prl} may enable an analysis of the time--evolution of this system more closely matched to the experiments than we have been able to perform here. This would be very interesting to probe in the future.

\section{Conclusion}

We have performed a detailed analysis of the string fluid structure in an anisotropic system of dipolar colloids and found reasonable agreement between our experiments and computer simulation data across a wide range of interactions tuned with the electric field. We found both bond--order parameter analysis of strings and the three--body correlation function $g_3$ to be suitable to quantify the degree of string formation in dipolar colloids but with $g_3$ offering more detailed information and can be used as a form of ``colloidal finger-print''. Using the topological cluster classification, we find that our experiments and simulations broadly agree with expectations from minimum energy clusters of a dipolar-Lennard-Jones system~\cite{skipper2024}. That is to say, structural transformations predicted at zero temperature for a Lennard--Jones--Dipolar system are rather effective in their prediction of higher--order structure in the nearly--hard sphere--dipolar experiments and simulations. At high field strength, the cluster population in both our experiments and simulations is dominated by the minimum energy clusters for all sizes $8 \leq m \leq 12$. Finally, not only can we identify clusters relevant to the dipolar system but also to investigate their orientation with respect to field strength.

\setcounter{figure}{0}
\renewcommand\thefigure{A\arabic{figure}} 

\begin{figure}
\begin{center}
\includegraphics[width=65 mm]{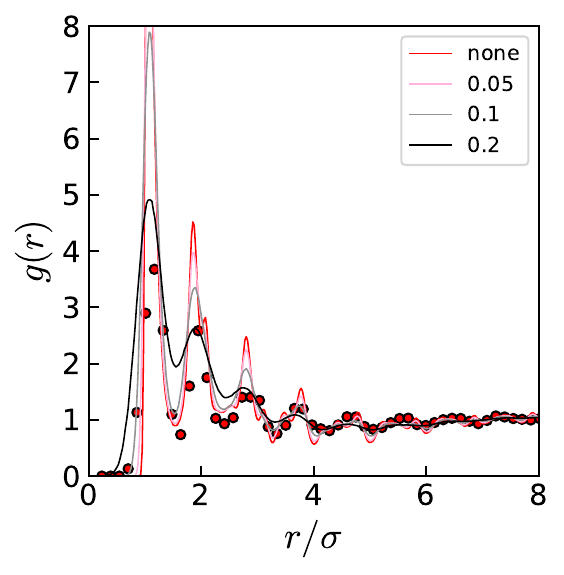}
\end{center}
\caption{The effects of adding errors to the coordinates on the radial distribution function $g(r)$. Shown is the data for a volume fraction $\phi=0.1$ and field strength $\gamma_\mathrm{eff}=50$. Experimental data (circles) and data without errors added to the simulated coordinates (red line) are reproduced from Fig.~\ref{figRadial}(a). Errors are added to the simulated coordinates using a Gaussian with standard deviation 0.05 (pink line) 0.1 (grey line) and 0.2 (black line). 
}
\label{sFigGError}
\end{figure}

\section*{Acknowledgements}
XW and FCM thank European Research Council (ERC) for funding an Advanced Grant for the DYNAMIN project, grant agreement No. 788968.
CPR acknowledges the Agence Nationale de Recherche for grant DiViNew.
We thank the KITP for hospitality while some of this work was done.
YY gratefully acknowledges the China Scholarship Council. 
FJM was supported by a studentship provided by the Bristol Centre for Functional Nanomaterials (EPSRC Grant No. EP/L016648/1). The authors would like to thank Mark Miller, Josh Robinson and Peter Crowther for the help in building the TCC software. The authors would also like to thank Mike Allen and Didi Derks for  helpful discussions.  %The authors are grateful 

%\bibliography{bulletproof} 

\end{document}